\documentclass[aps,prd,showpacs,twocolumn,letterpaper,preprintnumbers,nofootinbib]{revtex4-1}
\pdfoutput=1
\usepackage{amsmath,latexsym,amssymb,graphicx,enumerate}
\usepackage{slashed}
\usepackage{color}
\usepackage{cancel}
\usepackage{hyperref}
\usepackage{url}
\newcommand{\boldgreek}[1]{{\mbox{\boldmath{$#1$}}}}

\begin{document}
\title{Long-lived Light Mediator to Dark Matter and Primordial Small Scale Spectrum}
\author{Yue Zhang \vspace{0.1in}}
\affiliation{Walter Burke Institute for Theoretical Physics,\\
California Institute of Technology, Pasadena, CA 91125 \vspace{0.1in}\\
{\tt yuezhang@theory.caltech.edu}}

\begin{abstract}
We calculate the early universe evolution of perturbations in the dark matter energy density in the context of simple dark sector models containing a GeV scale light mediator. We consider the case that the mediator is long-lived, with lifetime up to a second, and before decaying it temporarily dominates the energy density of the universe. We show that for primordial perturbations that enter the horizon around this period, the interplay between linear growth during matter domination and collisional damping can generically lead to a sharp peak in the spectrum of dark matter density perturbation. As a result, the population of the smallest DM halos gets enhanced. Possible implications of this scenario are discussed.
\end{abstract}

\preprint{CALT-TH-2015-008}


\maketitle

\section{Introduction}

An important question for the field of particle cosmology is to reveal the nature of dark matter (DM).
There has been compelling evidence for the existence of DM, most of which are from gravitational effects.
There are also various ongoing experiments aiming to discover the DM through possible interactions other than gravity.
On the theory side, there are well-motivated frameworks like supersymmetry that can accommodate DM candidates as weakly interacting particles. An alternative approach to the theory of DM is using minimality as guiding principle.
The focus of this work is a class of simple models in which the DM candidate talks to the Standard Model (SM) sector
through the exchange of a light mediator. This setup has been proposed and explored for several different 
motivations~\cite{Pospelov:2007mp, ArkaniHamed:2008qn, Kaplan:2009ag, An:2009vq, Kaplan:2009de, Wise:2014jva}. With a light mediator it is possible to suppress the direct detection constraints on DM, but still have prospects for other indirect searches. The simple interaction between the light mediator and DM can offer rich dynamics in the dark sector~\cite{Wise:2014ola, Feng:2009mn, Buckley:2009in, Tulin:2013teo}.

In this note, we investigate the light mediator phenomenology and cosmology in the context of two simple models, where the mediator is either a vector or scalar boson. Its interaction with the SM sector is via photon kinetic mixing in the vector mediator case, and the Higgs portal in the scalar case.
We are interested in the mass range of the mediator at GeV scale, and lifetime as long as a second.
This is often a less studied region of parameter space because the portal interaction is so weak that no current experiment is sensitive to it. The aim of this work is to fill this gap by studying the consequences of the mediator's longevity and possible testable implications.
We note that the mediator is long-lived because of the tiny portal coupling between the dark and SM sectors---this also implies that the two sectors never reach equilibrium, but evolve with their own temperatures. 
There could be an interesting impact on the cosmology, if the light mediator becomes non-relativistic and temporarily dominates the energy density of the universe before it decays and releases entropy.
We calculate the evolution of density perturbations of DM during the mediator dominance 
and the imprint of this period on the primordial power spectrum.
Because the mediator has become matter-like when it begins to dominate the universe, 
the DM density perturbation could grow linearly. 
A competing effect is the collisional damping from the DM-mediator scattering.
We find that for perturbations entering the horizon during the mediator domination era, the linear growth effect wins over damping.
We show a generic prediction of such long-lived mediator model is a sharp peak in the transfer function right before the exponential cutoff. This corresponds to a characteristic length scale of order $10^{-4}$\,pc, that is, size of the minimal DM halos.
The peak implies these mini halos are to be more populated than in the usual cold DM case.

In the next section, we study the simplest dark sector models with a light vector/scalar mediator. 
We summarize the current experimental constraints on the
light mediator mass and its portal coupling to SM. 
The allowed region where the mediator can be long-lived and dominate the universe is highlighted.
In section III, we do the calculation on the evolution of density perturbations in the early universe, 
paying attention to the modes entering horizon around the light mediator dominated era.
We obtain the relative transfer function by comparing our results with those of the usual cold DM.
In section IV, we estimate the characteristic mass and size of the minimal DM halos and
discuss possible testable predictions. 

\begin{figure}[h]
\includegraphics[width=0.94\columnwidth]{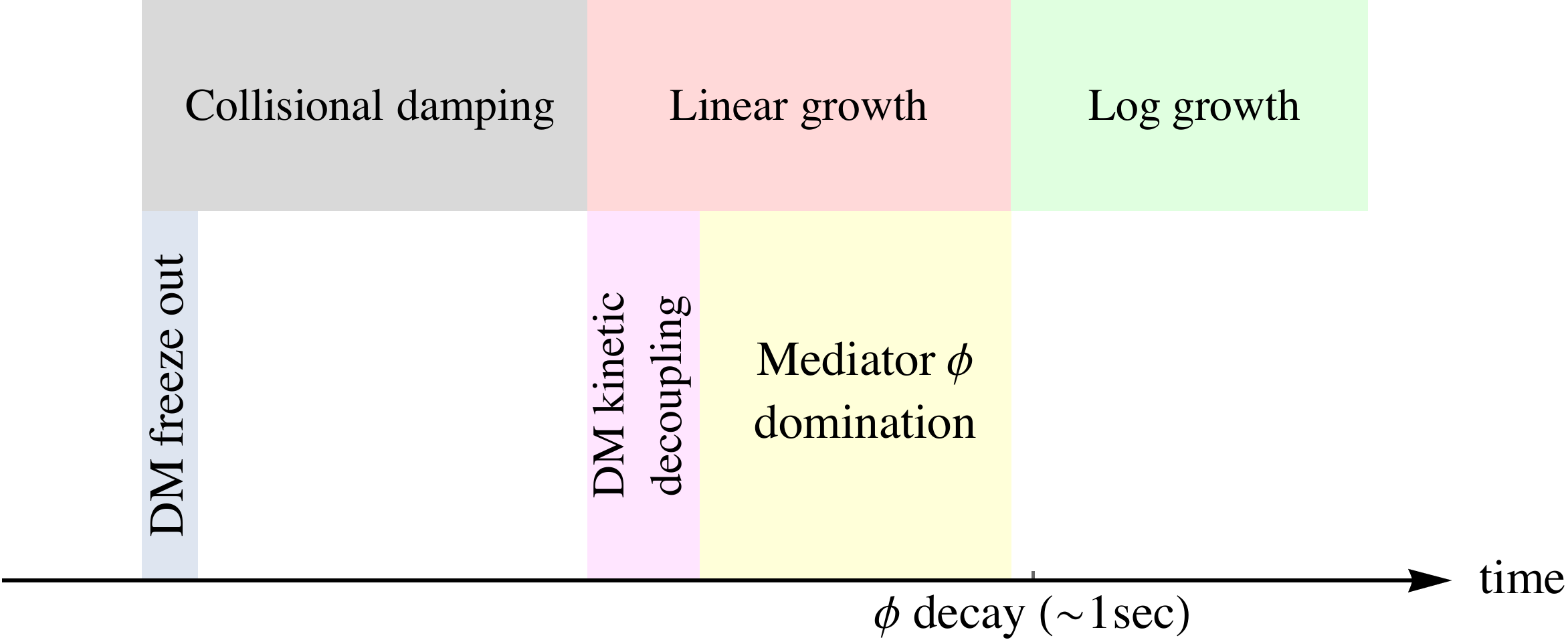}
\caption{History of dark sector and perturbation growth.}\label{1}
\end{figure}

\section{Simple dark sector models\\ with a light mediator}

To set the stage, we sketch in Fig.~\ref{1} the major events happening to the dark sector and their time scales relevant for this study.

\begin{figure*}[t!]
\centerline{\includegraphics[width=1.0\columnwidth]{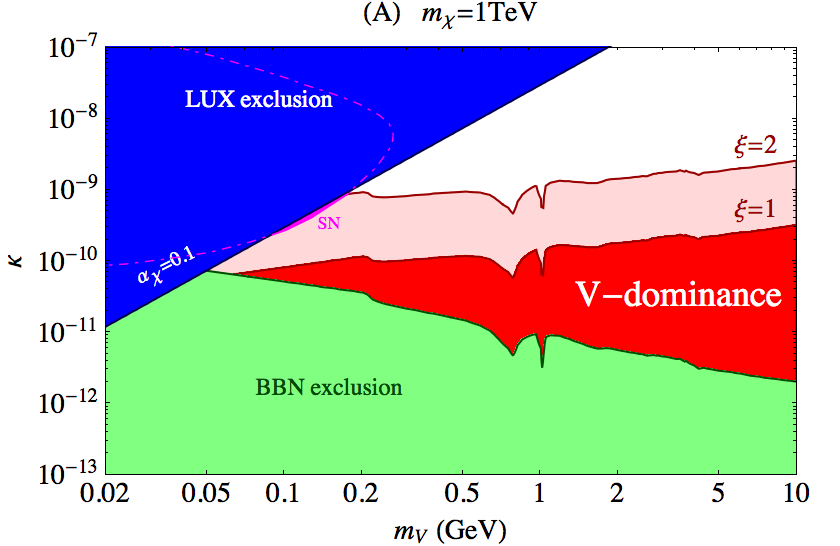}
\includegraphics[width=1.0\columnwidth]{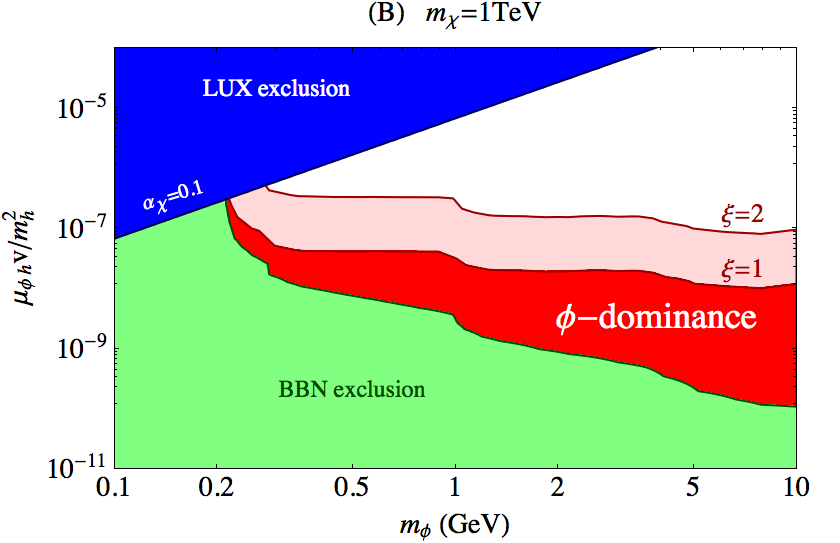}}
\vspace{0.3cm}
\centerline{\includegraphics[width=1.0\columnwidth]{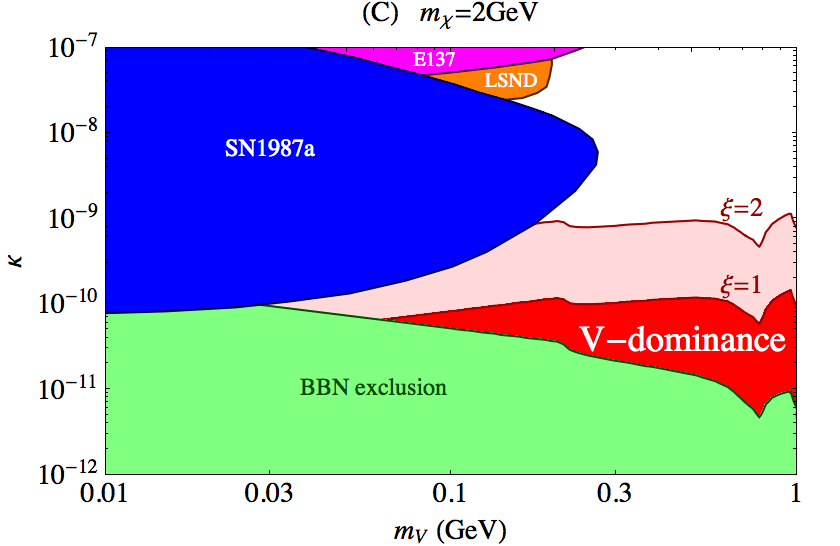}
\includegraphics[width=1.0\columnwidth]{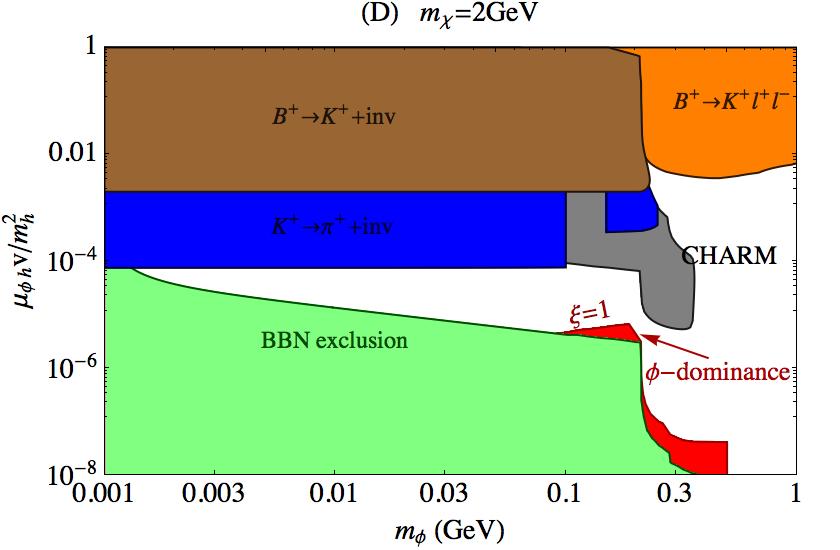}}
\caption{Experimental constraints on dark sector models with light mediator, shown in the parameter space of photon/Higgs portal versus mediator mass. We want to highlight the red region where the mediator could dominate the universe before it decays.}\label{bound}
\end{figure*}

\medskip
\noindent{\it\bfseries Vector mediator case}\\
The first dark sector model we examine contains a massive vector mediator $V_\mu$,
\begin{eqnarray}\label{Lvector}
\mathcal{L} \!&=&\! \mathcal{L}_{\rm SM} + \bar\chi i\gamma^\mu(\partial_\mu - i g_\chi V_\mu) \chi - m_\chi \bar \chi \chi \nonumber \\
\!&&\! -\frac{1}{4} V_{\mu\nu} V^{\mu\nu} + m^2_{V} V_\mu V^\mu - \frac{\kappa}{2} F_{\mu\nu} V^{\mu\nu} \ .
\end{eqnarray}
The DM field $\chi$ is a Dirac fermion and $V_\mu$ couples to the DM vector current. Both are SM gauge singlets. 
A kinetic mixing (the $\kappa$ term) between $V$ and the usual photon is introduced. 

In the early universe, the DM $\chi$ and the mediator $V_\mu$ are in thermal equilibrium through the gauge interaction
$g_\chi$. We focus on the case of light mediator, $m_\chi\gg m_V$. In this case, the DM freeze out is controlled by the annihilation $\chi\bar \chi \to VV$. To give the correct relic density to symmetric DM, the thermal annihilation cross section should satisfy 
\begin{eqnarray}\label{vectoranni}
\langle\sigma v\rangle_{anni} = \frac{\pi \alpha_\chi^2}{m_\chi^2} = \frac{3\times10^{-26} {\rm cm}^3/{\rm s}}{\mathcal{S}} \ ,
\end{eqnarray}
where $\mathcal{S}$ is the DM number density dilution factor due to any entropy production after the freeze out.
For asymmetric DM, the relic density is given the primordial asymmetry in $\chi$ and $\bar\chi$ number, and the annihilation cross section has to be larger than the thermal value.

After the DM freeze out, the light mediator $V_\mu$ takes over all the entropy in the dark sector. 
We assume the dark sector has its own temperature $T'$, and define its ratio to the visible photon temperature, $\xi\equiv T'/T_\gamma$, at very high temperature. After DM annihilation, the comoving $V_\mu$ number is
\begin{eqnarray}\label{nv}
{n_V}(T_\gamma) = \frac{13\zeta(3)\xi^3}{6\pi^2} \frac{g_{*s}(T_{cd})}{g_{*s}(m_t)}T_\gamma^3 \ ,
\end{eqnarray}
where $g_{*s}(T)$ counts the relativistic SM degree of freedom contributing to entropy density.
For this assumption to be valid, the interactions between $V_\mu$ and the SM particles must be decoupled
throughout the history of universe. 
Because the (kinetic) mixing $\kappa$ is the only portal that communicate the two sectors,
it is sufficient to require that the inverse decay rate $f\bar f \to V_\mu$ is less than the Hubble parameter at temperature of universe equal to $m_V$. \footnote{Although the other scattering processes such as $q\bar q \to V g$ may have larger rates at $T>m_V$ because $\Gamma\sim T$, the relevant quantity here is the ratio to the Hubble rate, $\Gamma/H$. 
Because $H\sim T^2$, these constraints are always less important than inverse decay.
A key ingredient for this to be valid is that the photon has a universal coupling constant to all fermions.
This is not true for the Higgs portal discussed in the next model.
}
We find for $m_\phi\leq10\,$GeV it requires $\kappa\lesssim10^{-7}$. 
As shown below, this is well satisfied if the light mediator is long-lived enough to dominate the universe.

There are further phenomenological constraints on $\kappa$ and $m_V$.
In order to be consistent with standard big-bang nucleosynthesis, a commonly used requirement is to make $V_\mu$ decay before 1 second or so. This gives a lower bound on $\kappa$. \footnote{If there is an additional very light scalar or fermion in the dark sector which the mediator can decay into, the constraint on $\kappa$ could be evaded.
This could lead to extra contribution to effective number of neutrinos, $N_{\rm eff}$~\cite{Weinberg:2013kea, Bertoni:2014mva}.
In this work, we focus on the minimal model and do not consider such extensions.} 
We use the numerical result on the decay rate in~\cite{Batell:2009yf} including the hadronic channels.
On the other hand, $\kappa$ also controls the direct detection cross section of DM $\chi$,
\begin{eqnarray}
\sigma_{\rm SI} \simeq \frac{16 \pi \kappa^2 \alpha_\chi \alpha \mu_{\chi N}^2}{m_\phi^4} \ ,
\end{eqnarray}
where $\alpha_\chi = g_\chi^2/(4\pi)$.
For weak scale DM, the current LUX~\cite{Akerib:2013tjd} experiment places the strongest upper limit on $\kappa$.
In order to satisfy both constraints, the mediator $V_\mu$ should be heavier than $\sim$\,50\,MeV. \footnote{The momentum exchange for DM scattering on xenon nucleus is roughly 50\,MeV. Thus for mediator lighter than this value, the cross section will get saturated by the momentum exchange, and the bound could be slightly weakened~\cite{Kaplinghat:2013yxa}. We find this only slightly affects the lower bound on the mediator mass.}
This is shown in the panel (A) of Fig.~\ref{bound}. 
We also notice that with a weak scale DM, the direct detection constraint on $\kappa-m_V$ parameter space is stronger than many of the limits on models with dark photon only.
On the other hand, if the DM is lighter than around $4\,$GeV, the direct detection limits from LUX and CDMSlite~\cite{Agnese:2013jaa} (which covers the lower mass region down to $m_\chi =4$\,GeV) are both evaded. 
However, as shown in panel (C) of Fig.~\ref{bound}, the other constraints from, {\it e.g.}, Supernova 1987a, are also very strong~\cite{Bjorken:2009mm, Dent:2012mx, Essig:2013lka} (for an exception see~\cite{Zhang:2014wra})---not much parameter space could be opened in the lower $m_V$ region.

We want to highlight a less constrained region where $10^{-12}\lesssim\kappa\lesssim10^{-9}$ and $m_V$ is at GeV scale. In this case, the $V$ particle is long-lived enough and becomes non-relativistic before it decays; 
the total energy density of the universe could experience a temporary period of $V$-matter domination.
To find the minimum lifetime of $V_\mu$ for this to happen, one can compare its energy density with that of the visible radiation species,
\begin{equation}
\begin{aligned}
\rho_{V}(T_\gamma) &= m_V {n_V}(T_\gamma) e^{- \frac{\Gamma_V}{2 H(T_\gamma)}} \ , \\
\rho_r(T_\gamma) &= \frac{\pi^2}{30} g_*(T_\gamma) T_\gamma^4 \left( \frac{g_{*s}(T_\gamma)}{g_{*s}(m_t)} \right)^{4/3} \ .
\end{aligned}
\end{equation}
Here $g_{*}(T)$ counts the relativistic SM degree of freedom contributing to energy density. 
After $T'<m_V$, the mediator becomes non-relativistic, and $\rho_V$ scales as $T_\gamma^{-3}$, 
while for radiation $\rho_r$ scales as $T_{\gamma}^{-4}$ up to the changes in $g_{*s}$.
In order for $\rho_{V}$ to be ever larger than $\rho_r$, we find the lower bound on the lifetime,
\begin{eqnarray}
\tau_{V} \gtrsim 0.004 {\xi}^{-6} \left( \frac{m_V}{1\,\rm GeV} \right)^{-2} \, {\rm sec} \ .
\end{eqnarray}
Note $0.004$ second corresponds to around a temperature of the universe at 15\,MeV.


\medskip
\noindent{\it\bfseries Scalar mediator case}\\
The Lagrangian for a dark sector with scalar mediator is
\begin{eqnarray}\label{Lscalar}
\mathcal{L} \!&=&\! \mathcal{L}_{\rm SM} + \bar \chi i\!\! \not \!\partial \chi - m_\chi \bar \chi \chi + \frac{1}{2} \partial_\mu \phi \partial^\mu \phi - m_\phi^2 \phi^2 + y_\chi \bar \chi \chi \phi \nonumber \\
\!&&\!+ \mu_{\phi h} (H^\dagger H - v^2/2) \phi + \lambda_{\phi h} H^\dagger H \phi^2 + \lambda_\phi \phi^4 \ ,
\end{eqnarray}
where the DM $\chi$ is a Dirac fermion, and the mediator $\phi$ is a real scalar, and both are SM singlets.

In early universe $\chi$ and $\phi$ are in thermal equilibrium via their Yukawa coupling $y_\chi$.
For $m_\chi\gg m_\phi$, the DM relic density is controlled by the process $\chi \bar \chi \to \phi \phi$. 
For symmetric DM, correct thermal relic density requires the annihilation cross section satisfy
\begin{eqnarray}\label{scalaranni}
\langle\sigma v\rangle_{anni} = \frac{9\pi \alpha_\chi^2 T_{cd}}{2 m_\chi^3} = \frac{3\times10^{-26} {\rm cm}^3/{\rm s}}{\mathcal{S}} \ .
\end{eqnarray}
This is a $p$-wave annihilation and $T_{cd} \approx m_\chi/26$ is the chemical decoupling temperature.
Again we assume the dark sector has its own temperature $T'$. After DM freeze out the comoving $\phi$ number is given by
\begin{eqnarray}\label{nphi}
{n_\phi} = \frac{9\zeta(3)\xi^3}{2\pi^2} \frac{g_{*s}(T_{cd})}{g_{*s}(m_t)}T_\gamma^3 \ .
\end{eqnarray}

In this model, there are two ways that $\phi$ could communicate to SM.
First, if the coefficient of cross quartic term $\lambda_{\phi h}$ is large, $\phi$ will experience a similar freeze out via $s$-channel Higgs exchange. 
Second, the $\mu_{\phi h}$ term in the Lagrangian induces a mixing between $\phi$ and the Higgs boson and is responsible for $\phi$ to scatter and decay. 
Because of the Higgs portal, the $\phi$-top quark scattering is most important in bringing $\phi$ into equilibrium with SM than inverse decay. 
As said, we are primarily interested in a similar parameter space as the vector mediator case, where $\lambda_{\phi h}$, $\mu_{\phi h}$ are small enough so that they never bring $\phi$ into equilibrium with the SM sector.
This requires, $\lambda_{\phi h}\lesssim10^{-7}$ and $\mu_{\phi h} v/m_h^2\lesssim 10^{-7}$, respectively.

The BBN and direct detection could further constrain $\mu_{\phi h}$ and $m_{\phi}$.
For the $\phi$ decay rate we use a result including the hadronic channels~\cite{Voloshin:1985tc, Clarke:2013aya}.
The direct detection cross section of DM $\chi$ is 
\begin{eqnarray}
\sigma_{\rm SI} \simeq \frac{4 \alpha_\chi f^2 m_N^2 \mu_{\chi N}^2 \mu_{\phi h}^2}{m_h^4 m_\phi^4} \ ,
\end{eqnarray}
where $\alpha_\chi = y_\chi^2/(4\pi)$.
It was noted in~\cite{Wise:2014jva} that if DM is heavier than $\sim$\,5GeV, $\phi$ must weigh more than twice of the muon mass, $m_\phi \gtrsim 210\,$MeV. This can be seen from panel (B) of Fig.~\ref{bound}.

Because there are fewer constraints on a light scalar mediator than a vector one, 
for DM mass at GeV scale, we find it possible to lift the direct detection constraint, and make $\phi$ as light as MeV. There is still an upper bound on $\phi$-Higgs mixing from the low energy $K^+\to\pi^++\!\!\not{\!\!E_T}$, $K^+\to\pi^+\ell^+\ell^-$, $B^+\to K^+\ell^+\ell^-$ processes~\cite{Anchordoqui:2013bfa}, and the fix-target CHARM experiment~\cite{Bezrukov:2009yw}, as shown in panel (D) of Fig.~\ref{bound}. 
In together with the decay before BBN requirement, there is a narrow window still allowed for the parameter
$10^{-5}\lesssim\mu_{\phi h} v/m_h^2\lesssim10^{-4}$.
However, in this window, $\phi$ will always be thermalized with SM particles at high temperature, which fixes $\xi=1$.

Similar to the vector mediator case,
we are interested in the possibility that $\phi$ temporarily dominates the total energy density of the universe before it decays. Using (\ref{nphi}) we derive the minimal lifetime of $\phi$ for this to happen,
\begin{eqnarray}
\tau_{\phi} \gtrsim 0.01 {\xi}^{-6} \left( \frac{m_\phi}{1\,\rm GeV} \right)^{-2} \, {\rm sec} \ .
\end{eqnarray}
This happens if $10^{-10}\lesssim\mu_{\phi h} v/m_h^2\lesssim10^{-8}$, and is highlighted in the red region of Fig.~\ref{bound} (B). 

To close this section, we have summarized the experimental constraints on two simple models of dark sector with light mediator, $V$ or $\phi$, and showed the regions in the parameter space that allow the mediator to temporarily dominate the universe as matter. The next few sections are devoted to the study of the implication of this possible era on the DM primordial spectrum. 

\section{Light mediator dominance\\ and perturbations}

To calculate the perturbation to energy densities in early universe, we define the metric in the
conformal Newton gauge
\begin{eqnarray}
ds^2 = - (1+2\Psi) dt^2 + (1+2\Phi)a^2 \delta_{ij} dx^i dx^j \ .
\end{eqnarray}
The perturbations satisfy $\Psi=-\Phi$ in the absence of quadrupole moment.

There are three species in the universe\footnote{In this section we will proceed the discussion a with scalar light mediator $\phi$. The results for vector mediator case are very similar.}, DM $\chi$, light mediator $\phi$ and SM radiation $r$.
The radiation includes all relativistic SM particles which tightly couple to each other. For each species, the stress-energy tensor and the perturbations are
\begin{eqnarray}
T^{\mu\nu} = (1+w) \rho (1+\delta) u^\mu u^\nu + w \rho (1+\delta) g^{\mu\nu} \ ,
\end{eqnarray}
where $w=p/\rho$, and $u^\mu = (1-\Psi, {\bf v})$. 
The next-to-leading order quantities are $\delta$ and ${\bf v}$.

Using the continuity of $T_{\mu\nu}$, the decay of $\phi$ into usual radiation with DM as a spectator can be described as~\cite{Scherrer:1984fd}
\begin{eqnarray}\label{rhos}
&&\dot \rho_\chi + 3 H \rho_\chi=0 \ , \nonumber \\
&&\dot \rho_\phi + 3 (1+w_\phi) H \rho_\phi= - \Gamma_\phi \rho_\phi \ ,  \\
&&\dot \rho_r + 4 H \rho_\chi=\Gamma_\phi \rho_\phi \ .\nonumber
\end{eqnarray}
where an over dot means $d/dt$. 
The Hubble expansion rate is
\begin{eqnarray}
H^2 = \frac{8\pi G}{3}  \left( \rho_\chi + \rho_\phi + \rho_r \right) \ .
\end{eqnarray}
We include the process where the light mediator $\phi$ changes from radiation-like ($w_\phi=1/3$) to matter-like ($w_\phi\to0$) as the temperature drops.
As assumed throughout this work, after DM freeze out the total number of $\chi$ no longer change, and so does that of $\phi$ until it begins to decay to radiation species of the SM sector.

\begin{figure}[t!]
\includegraphics[width=1\columnwidth]{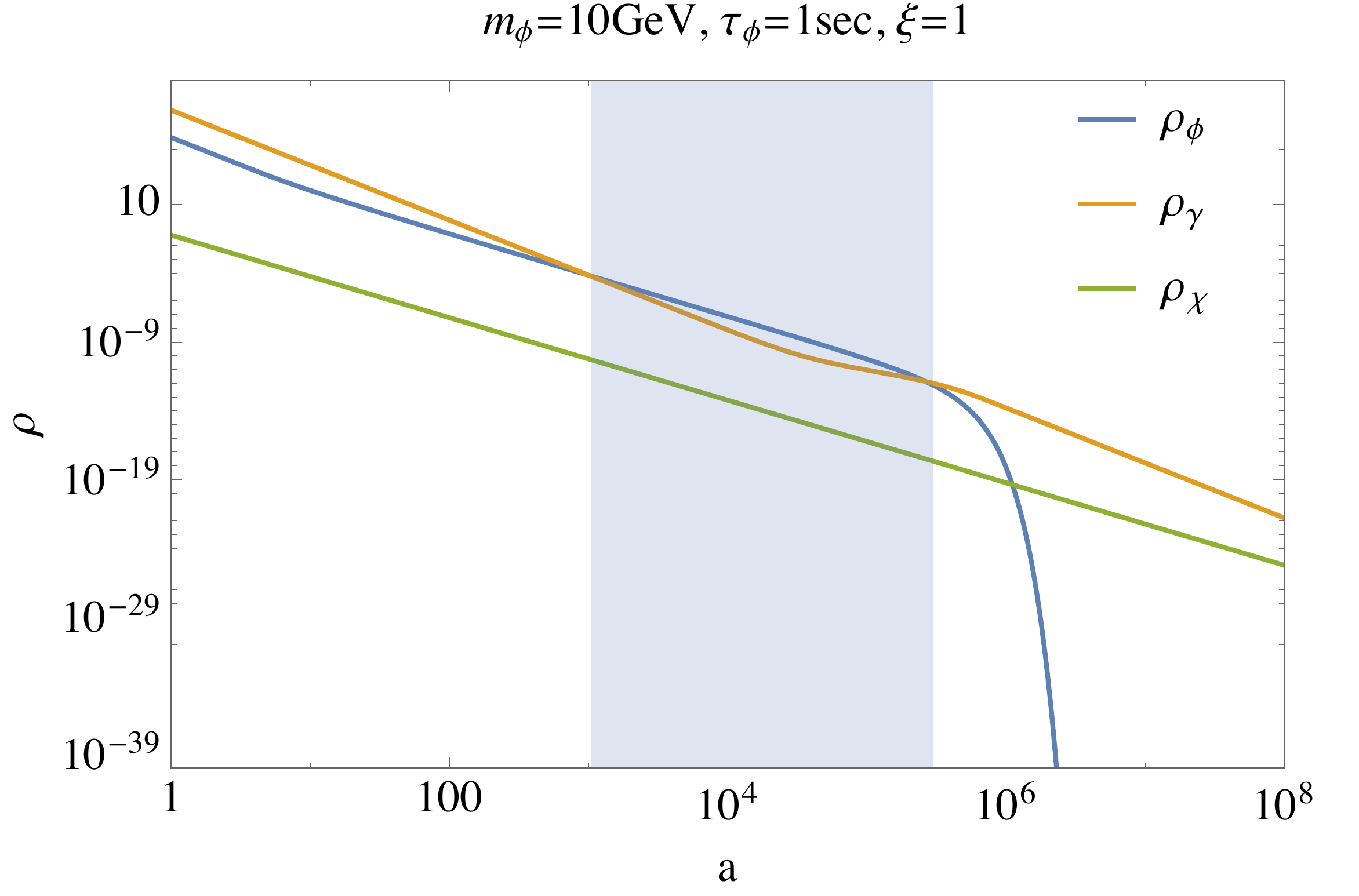}
\caption{Evolution of the energy densities of DM $\chi$, light mediator $\phi$ and SM radiation $r$. In the shaded region, the universe is $\phi$ dominated.}\label{rho_t}
\end{figure}

\begin{figure*}[t!]
\centerline{
\includegraphics[width=2.0\columnwidth]{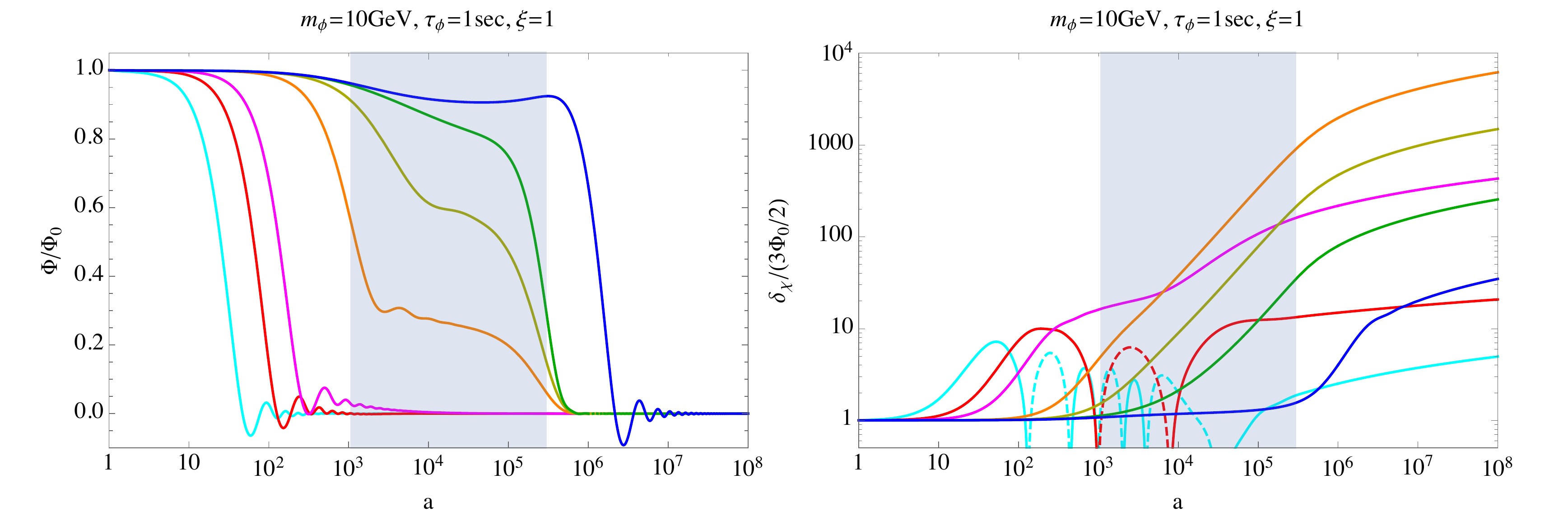}
}
\caption{Evolution of primordial perturbations of different wavelengths that enter the horizon before, during and after the $\phi$ domination regime. Different colors correspond to wave numbers $k/k_{eq}=10^{9.5}$ (cyan), $10^{9.1}$ (red), $10^{8.8}$ (magenta), $10^8$ (orange), $10^{7.5}$ (yellow), $10^7$ (green) and $10^6$ (blue), respectively. The dashed part of curves means sign change in the perturbation due to acoustic oscillation.}\label{delta_t}
\end{figure*}

We present the results using a set of sample input parameters,
$m_\phi=10\,$GeV $\tau_\phi=1\,$sec, $\xi=1$, and $m_\chi=1\,$TeV. We also choose the initial temperature to be $T_{ini}=m_\chi/26\gg m_\phi$ right after freeze out, and define the Hubble radius at this time to be $a_0\equiv1$.
As a good approximation, $T_\gamma = T_\phi = T_{ini}$ at the time $a_0$.
The initial conditions for (\ref{rhos}) are
\begin{eqnarray}
\rho_\chi (a_i) &=&  \mathcal{S} \times ({5.4\,\rm GeV}) \times \eta_b  \frac{2\pi^2}{45} g_{*s}(T_{ini}) T_{ini}^3 \ , \nonumber \\
\rho_\phi (a_i) &=& \frac{\pi^2}{30} T_{ini}^4 \ ,  \ \ \ \ 
\rho_r (a_i) = \frac{\pi^2}{30} g_*(T_{ini}) T_{ini}^4 \ .\nonumber
\end{eqnarray}
where $\eta_b=n_b/s=0.9\times10^{-10}$ is the baryon-to-entropy ratio today, and the dilution factor $\mathcal{S}$ due to entropy release from $\phi$ decay is fixed a posteriori from solving Eqs.~(\ref{rhos}),
\begin{eqnarray}
\mathcal{S} = \left( \frac{\rho_\chi (a_i)}{\rho_\chi (a_f)} \right) \left( \frac{\rho_r (a_f)}{\rho_r (a_i)} \right)^{3/4} \ ,
\end{eqnarray}
where $a_f$ corresponds to an epoch well after $\phi$ decay.

Fig.~\ref{rho_t} shows the evolution of the $\rho$'s.
During the period $10^3\lesssim a\lesssim 3\times10^5$, the energy density of $\phi$ beats that of radiation and dominates the total energy in the universe. This happens until the $\phi$ particles decay away, 
and dump a significant amount entropy into radiation.
With the above set of parameters, $\mathcal{S} \simeq 60$, and matter-radiation equality happens at $a_{eq}=2.8\times10^{11}$.

The equations for the evolution of perturbations in the fluid densities and in the metric are
\begin{eqnarray}\label{EqPert}
&&\dot \delta_\chi + \frac{\theta_\chi}{a} + 3 \dot \Phi=0 \ , \\
&&\dot \theta_\chi + H \theta_\chi + \frac{k^2}{a} \Phi = c_s^2 \frac{k^2}{a} \delta_\chi + \tau_c^{-1} (\theta_\phi - \theta_\chi) \ , \nonumber \\
&&\dot \delta_\phi + (1+w_\phi)\frac{\theta_\phi}{a} + 3 (1+w_\phi) \dot \Phi = \Gamma_\phi \Phi \ , \nonumber \\
&&\dot \theta_\phi + (1-3w_\phi) H \theta_\phi + \frac{k^2}{a} \Phi - \frac{w_\phi}{1+w_\phi} \frac{k^2}{a}  \delta_\phi=\frac{w_\phi}{1+w_\phi} \Gamma_\phi \theta_\phi , \nonumber \\
&&\dot \delta_r + \frac{4\theta_r}{3a} + 4 \dot \Phi= \frac{\rho_\phi}{\rho_r}\Gamma_\phi \left(\delta_\phi - \delta_r - \Phi \right) \ , \nonumber \\
&&\dot \theta_r - \frac{k^2}{4a} \delta_r + \frac{k^2}{a} \Phi= \frac{\rho_\phi}{\rho_r}\Gamma_\phi \left( \frac{3}{4} \theta_\phi - \theta_r \right) \ , \nonumber \\
&&\dot \Phi + \left( H + \frac{k^2}{3 H a^2} \right) \Phi = \frac{4\pi G}{3H} \left( \rho_\chi \delta_\chi + \rho_\phi \delta_\phi + \rho_r \delta _r \right) \ , \nonumber
\end{eqnarray}
where $\theta\equiv a {\boldgreek \nabla} \cdot {\bf v}$ for each species. We derived these equations following the notations in~\cite{Ma:1995ey, Erickcek:2011us, Fan:2014zua, Bertschinger:2006nq}. Notice again the over dot means $d/dt$ here. 
The rate $\tau_c^{-1}$ is related to the scattering process $\phi\chi\to\phi\chi$ which keeps the DM in kinetic equilibrium with $\phi$, and it is given by
\begin{eqnarray}
\tau_c^{-1} \simeq n_\phi v_\phi \left(\frac{4\pi \alpha_\chi^2}{3m_\chi^2}\right) \left( \sqrt{\frac{3}{2}} \frac{T}{m_\chi}\right) \ ,
\end{eqnarray}
where the first bracket is the $\phi\chi\to\phi\chi$ cross section\footnote{For the vector mediator model, the $V\chi\to V\chi$ cross section is that of Thomson scattering, ${8\pi \alpha_\chi^2}/({3m_\chi^2})$.} at $T\ll m_\chi$, and the second bracket represents the ratio of collision time to the thermal relaxation time~\cite{Hofmann:2001bi}.
The relative velocity $v_\phi=1$ when $T\gtrsim m_\phi$, and $v_\phi=\sqrt{3T/m_\phi}$ at low temperature.
For $\alpha_\chi=0.1$ and $m_\chi=1\,$TeV, this rate falls below Hubble at $a_{kd}\simeq10^4$ (around $T\sim10\,$MeV).
With this input, the sound speed $c_s$ of the the $\chi$ fluid is, 
$c_s^2=\delta p_\chi/\delta\rho_\chi=2T_\phi/(3m_\chi)$ before kinetic decoupling ($a \lesssim a_{kd}$), and more suppressed for $a \gtrsim a_{kd}$.
The back reaction of the scattering on the $\theta_\phi$ equation has been neglected because $\phi$ is much more populated than $\chi$.

The primordial scalar perturbations does not evolve until entering the horizon. The initial conditions for (\ref{EqPert}) can be obtained by simplifying these equations for super-horizon modes. At time $a_0$, the universe is still radiation dominated, thus,
\begin{eqnarray}
&&\delta_\chi (a_0)=\delta_\phi (a_0) = \frac{3}{4} \delta_r (a_0) = \frac{3}{2} \Phi_0 \ , \nonumber \\
&&\theta_\chi (a_0)=\theta_\phi (a_0)=\theta_r (a_0) = - \frac{k^2 \Phi_0}{2 H(a_0)} \ ,
\end{eqnarray}
where $\Phi_0$ is the primordial scalar perturbation in the metric.

We solve the set of equations in (\ref{EqPert}) for modes that enter the horizon before, during and after the $\phi$ domination regime. The results on $\Phi$ and $\delta_\chi$ are shown in Fig.~\ref{delta_t}.
From the $\Phi$ plot one could infer when a mode enters the horizon, while the $\delta_\chi$ plot shows the impact of
$\phi$-$\chi$ scattering and the era of $\phi$ matter domination on different scales.

\begin{figure}[t]
\includegraphics[width=1\columnwidth]{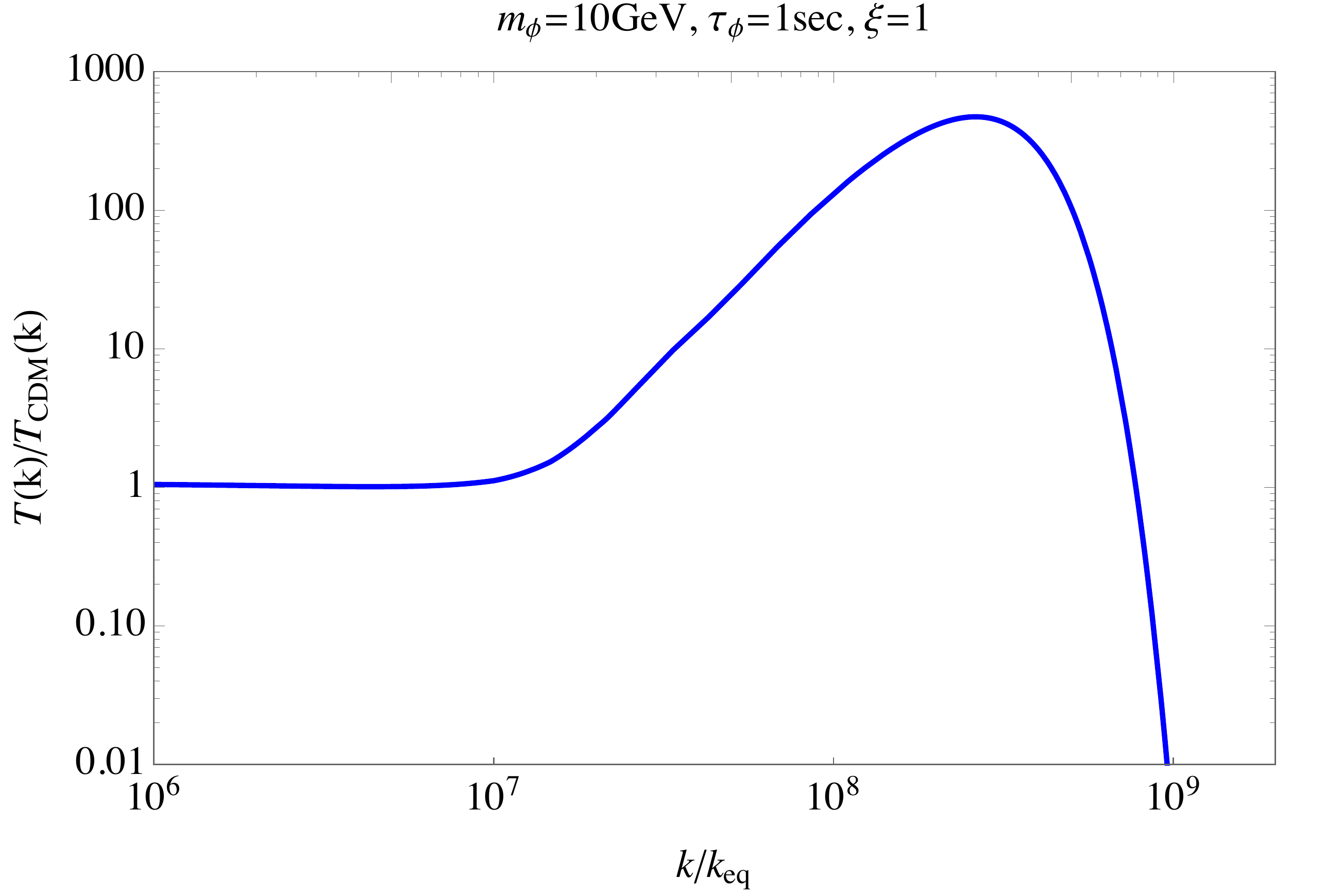}
\caption{The transfer function normalized to that of collision-less cold DM.
$k_{eq}$ corresponds to the mode that enters the horizon at matter-radiation equality.
}\label{tk}
\end{figure}

During the period when $\rho_\phi>\rho_r$, the universe is temporarily matter dominated,
and it is well known that in this period the density perturbations of $\phi$ itself and DM $\chi$ both grow linearly.
This feature is shown in Fig.~\ref{delta_t}, where $\phi$ domination depicted by the shaded region.
For modes entering the horizon in this period, $\Phi$ drops slowly allowing $\theta_\chi$ to grow, which then serves 
as the source for the linear growth of $\delta_\chi$.
After $\phi$ begins to decay and the universe returns to radiation dominated, these perturbations grow logarithmically. 
Because all $\phi$ particles eventually decay away, only the DM density perturbation $\delta_\chi$ matters for later structure formation at this scale, and it still remembers the $\phi$-domination history.
The enhancement in $\delta_\chi$ implies it will reach the nonlinear growth regime much earlier than the usual cold DM case.

At earlier time, the effect of $\phi$-$\chi$ scattering plays an important role. 
The scattering term $\tau_c^{-1}$ couples the $\phi$ fluid tightly to the DM $\chi$ fluid.
Because in this regime $\phi$ still carries non-negligible kinetic energy and has a larger sound speed, it drives the acoustic oscillation
and causes collisional damping in $\delta_\chi$ until the rate falls below $H$ at $a_{kd}$.
At very high wavenumber $k$, the resulting damping factor in the $\delta_\chi$ spectrum can be estimated as~\cite{Green:2005fa},
\begin{eqnarray}\label{damp}
D_{damp} (k) = \exp \left( - \frac{k^2}{k_{cut}^2} \right) \ .
\end{eqnarray}
where we find $k_{cut}$ corresponds to the mode entering the horizon at the moment of $\phi$ domination begins. 
For modes entering the horizon during $\phi$ dominated era, they are less affected by the collisional damping effects, although the kinetic decoupling happens afterwards. This can be understood because both $\phi$ and $\chi$ have become effectively non-relativistic, and their perturbations satisfy the same equation with the same initial condition. The momentum exchange term $\tau_c^{-1} (\theta_\phi - \theta_\chi)$ is also not important because
$\theta_\phi$ and $\theta_\chi$ evolve hand in hand with each other.

To obtain the transfer function $T(k)$, we evolve the DM density perturbation $\delta_\chi$ to the time of matter-radiation equality, and normalize the perturbations to those of a cold DM. The result is shown in Fig.~\ref{tk}. 
The relative transfer function $T(k)/T_{\rm CDM}(k)$ exhibits a peak at wavenumber $\sim$\,$10^{8}k_{eq}$ right before the exponential cutoff.
The enhancement factor in this case in the density perturbation can be as large as $100-1000$.

This peak-plus-cutoff feature is a unique prediction of dark sector models where DM couples to a long-lived light mediator.
This result also does not depend on whether the DM is symmetric or asymmetric.

\section{Possible testable implications}

From DM density perturbation spectrum one can calculate the DM structure formation at later time.
The peak before the cutoff in the primordial spectrum corresponds to a special scale of the smallest DM halos/clumps.
The peak right before the cutoff indicates that these smallest objects are most likely formed.
Here we estimate their properties.

From Fig.~\ref{tk}, we get the wavenumber at the peak in spectrum, $k_{peak} \sim 10^{8} k_{eq}$. 
The mass of the smallest objects today can be determined $k_{peak}$, evaluated using~\cite{Loeb:2005pm},
\begin{eqnarray}
M_{peak} = \frac{4\pi \Omega_{dm}^0 \rho_c^0}{3} \left( \frac{\pi}{k_{peak}} \right)^3 \simeq 10^{-6} M_{\odot} \ .
\end{eqnarray}
where we used today's DM relic density $\Omega^0_{dm}=0.26$, today's critical density $\rho_c^0=8.5\times10^{-30}\,{\rm g\, cm^{-3}}$, $k_{eq}=0.073\,{\rm Mpc^{-1}} \Omega^0_{DM} h^2$, $h=0.673$, and $M_\odot$ is the solar mass.

The size of these halos is given by the length scale of the mode $k_{cut}^{-1}$ at the time when the DM density perturbation $\delta_\chi(a_{nl})$ approaches to 1~\cite{Gurevich:1998qp},
\begin{eqnarray}
R_{peak} = \left( \frac{3M_{peak}}{4\pi \rho_{dm}(a_{nl})} \right)^{1/3}  \ .
\end{eqnarray}
Clearly $R_{peak}$ is more sensitive to the linear growth during $\phi$ domination. 
To get $a_{nl}$, we first assume scale invariance in the primordial scalar perturbation and fix the initial value of $\Phi_0\simeq 8\times10^{-6}$ from the Harrison-Zel'Dovich-Peebles spectrum in the large scale structure observation~\cite{dodelson}.
From this, we find for the mode corresponding to the peak of Fig.~\ref{tk}, $\delta_\chi(k_{peak})$ will reach nonlinear growth shortly after recombination, and $R_{peak}\sim 10^{14}\,$cm. The average DM density inside such mini halos is about $10^9$ times higher than the local DM density in the usual case.

Therefore, the first formed DM halos have a typical mass similar to that of the earth, and typical size similar to the earth-sun distance. After the formation, these halos may merge into larger objects or be destroyed by gravitational tidal forces.
A more precise calculation of the small scale structure formation and properties of such objects today calls for 
simulation and is beyond the scope of this work. 
Today, if the smallest halos still takes substantial fraction of total Milky Way galaxy mass, the average distance of two nearby such objects is $\sim 10^{16}\,$cm. Thus the solar system would run into such objects every 50-100 years and the direct detection rates will get enhanced periodically.
Another way to observe these clumps is indirect detection signals from, {\it e.g.}, DM annihilation from inside.
For asymmetric DM case, if the self-interaction allows DM bound states to exist and form, as discussed in the context of Yukawa bound states~\cite{Wise:2014jva} or the case of dark atom~\cite{Pearce:2015zca}, the indirect detection signal is also available. It could be enhanced if the bound state formation makes these clumps dissipate enough and further reduce their sizes.  There have also been suggestions that large primordial density perturbations could lead to micro-lensing effect 
if DM forms more compact objects~\cite{lensing}.

\section{Outlook}

To summarize, in the context of simple dark sector models containing a long-lived light mediator, we notice the mediator's longevity can be closely connected to an era of temporary matter domination. We have calculated the early universe evolution of dark matter density perturbations, 
especially during the mediator domination era when the perturbation could grow linearly. We showed this scenario can result in a sharp peak in the transfer function compared to that of collision-less cold dark matter.
This leaves us with more structures on small scales to look for, and could provide additional handle in testing light mediator models with very weak interactions to the standard model. More dedicated simulation of the small scale structure formation would help to obtain more concrete prediction in the properties of these mini dark matter halos.

{We note a temporary matter dominated universe before BBN has been discussed in various warm DM models, such as wino DM produced from late decaying moduli~\cite{Acharya:2009zt}, and keV sterile neutrino DM diluted by long-lived right handed neutrinos~\cite{Asaka:2006ek, Nemevsek:2012cd, Bezrukov:2009th}. In the moduli decay case, the free streaming of boosted wino always washes out most structures on small scales~\cite{Erickcek:2011us, Fan:2014zua}.  The free streaming nature of light sterile neutrino will result in the same suppression. In contrast, in the light mediator models discussed in the present work, the key for the primordial small scale structures to survive is that the long-lived mediator cannot decay to DM, and the DM itself remains cold.}

\section*{Acknowledgements.} 
We thank Sean Carroll, Clifford Cheung, Francis-Yan Cyr-Racine, Hooman Davoudiasl, Jonathan Feng, Michael Graesser, Hong-Jian He, Goran Senjanovic, Tim Tait, Mark Wise, Hai-bo Yu for useful conversations and comments on the draft.
This work is supported by the Gordon and Betty Moore Foundation through Grant \#776 to the Caltech Moore Center for Theoretical Cosmology and Physics, and by the DOE Grant DE-FG02-92ER40701, and also by a DOE Early Career Award under Grant No. DE-SC0010255.

\end{document}